\documentclass[11pt, aps, prd, tightenlines, nofootinbib, superscriptaddress, notitlepage]{revtex4-1}
\usepackage{xspace}
\usepackage[english]{babel}
\usepackage{amsfonts, amsmath, amsthm, amssymb, mathrsfs, bbm, stmaryrd}
\usepackage{graphicx}
\usepackage[left=2cm, right=2cm, top=2.5cm, bottom=2.5cm]{geometry}
\usepackage[colorlinks=true, citecolor=cyan, linkcolor=., urlcolor=.]{hyperref}
\usepackage{enumerate}
\usepackage[dvipsnames]{xcolor}
\usepackage{soul}

\newcommand{\abs}[1]{\left|#1\right|}
\newcommand{\bra}[1]{\langle #1 |}

\newcommand{\ket}[1]{|#1\rangle}
\newcommand{\norm}[1]{\left|\left|#1\right|\right|}
\newcommand{\mean}[1]{\left\langle #1 \right\rangle}
\newcommand{\Tr}{\text{Tr}}
\newcommand{\Pexp}{\text{$P$exp}}
\newcommand{\Pint}[1]{\mathcal{P}\!\!\!\!\!\!\int_{#1}}

\newcommand\scalemath[2]{\scalebox{#1}{\mbox{\ensuremath{\displaystyle #2}}}}

\renewenvironment{align}{\begin{equation}
\begin{aligned}}
{\end{aligned}
\end{equation}\par}

\begin{document}

\title{Loop representation and r-Fock measures for $SU(N)$ gauge theories}

\author{Mehdi Assanioussi}
\email[]{mehdi.assanioussi@fuw.edu.pl}
\affiliation{Faculty of Physics, University of Warsaw, Pasteura 5, 02-093 Warsaw, Poland.}

\author{Jerzy Lewandowski}
\email[]{jerzy.lewandowski@fuw.edu.pl}
\affiliation{Faculty of Physics, University of Warsaw, Pasteura 5, 02-093 Warsaw, Poland.}

\begin{abstract} 
In this article we continue the work on translating elements of the perturbative quantum field theory defined on Minkowski spacetime into the background independent framework of Loop Quantum Gravity. We present the construction of r-Fock measures for $SU(N)$ gauge theories and provide a relation between these new r-Fock measures and the difeomorphism invariant measure used in loop quantum gravity.
\end{abstract}

\maketitle

\section{Introduction}

The loop quantum gravity (LQG) program \cite{LQG0, LQG1, LQG2, LQG3, LQG4} defines a framework where gravity and the Standard Model matter fields are quantized in a background independent setting.
There has been many developments in this program and the resulting theory has very solid mathematical foundations. However several issues remain unresolved, in particular the recovery of the continuum limit of the theory. This issue can roughly be phrased in the question of how quantum field theory on a fixed spacetime arises from the quantum theory of gravity coupled to matter fields.

Understanding the continuum limit is necessary in order to be able to extract physical predictions from the fundamental theory. This is the reason why it becomes important to apprehend the link between the loop quantization and the Fock quantization in a simpler context, where gravity is in the semi-classical regime. This may help evaluating the consistency of the loop quantization with the well-established Standard Model, and may provide an intuition to construct the proper tools to treat the quantum regime of gravity. There has been several works in this direction, among which are the so called r-Fock representations.
In general terms, the r-Fock representations in loop quantum gravity can be defined as specific quantum representations of the algebras of holonomies associated to gauge fields on fixed spacetime. Introduced in \cite{Var1,Var2,Var3} for Abelian gauge fields and for a scalar field in \cite{AshLewSah} on Minkowski spacetime, they are intermediate representations which aim at connecting the standard Fock representation and the background independent loop representation. It is also argued that for finite accuracy measurements at distance scales much larger than a certain parameter characterizing the $r$-Fock representations, these representations are physically indistinguishable from the standard Fock representation.
They ultimately provide an embedding of Fock states into the Hilbert space of the loop representation, and hence a concrete representation of some (approximate) physical states in the loop quantum theory. In particular, they give rise to the so-called shadow states \cite{AshLew1}, which are projections of Fock states into separable Hilbert subspaces. These shadow states encode information about the fixed background geometry on which the matter field is propagating, however their properties remain largely unexplored and their relation to the quantum dynamics is entirely not understood. 

In \cite{Var1,Var2,Var3}, the r-Fock representation is defined as a representation of the standard holonomy-flux algebra of an Abelian gauge field on Minkowski spacetime, and the construction relies on the fact that the standard holonomy-flux algebra is isomorphic to an algebra of ``smeared holonomies-electric field'' algebra which admits a Fock quantization. Thanks to the existence of this algebra isomorphism, the Fock representation of the ``smeared holonomies-electric field'' algebra provides the r-Fock representation of the standard holonomy-flux algebra. Such isomorphism, however, does not exist for non-Abelian gauge theories. This fact was considered an obstruction for the generalization of the construction of r-Fock representations for non-Abelian gauge theories. Nevertheless, in \cite{AshLew1} the authors proposed the construction of an r-Fock measure for the non-Abelian case by generalizing an operator-form of the mapping between the natural difeomorphism invariant measure in LQG and the r-Fock measure for $U(1)$ gauge theory. While the resulting measure is indeed well-defined, it suffered some issues of gauge invariance and non-locality. In the present article, we propose a different method to construct an r-Fock measure for non-Abelian gauge theories, which follows a generalization of some of the steps of the construction in the Abelian case and avoids the need for the algebra isomorphism. Our motivation to generalize the notion of r-Fock measures stems from the perspective that the r-Fock representations could play a role in understanding the emergence of the continuum limit of loop quantum gravity and providing elements for the construction of a renormalization process for the theory.


The article is organized as follows. In section II we review the construction of the r-Fock representation for a $U(1)$ gauge theory. In section III, we develop our construction of r-Fock measures for $SU(N)$ gauge theories. We explicitly work out the details of the construction in the case of an $SU(2)$ gauge theory by first introducing the different classical algebras, then describing the Fock quantization of the smeared holonomies algebra, and eventually defining the r-Fock measure and establishing its relation with the natural difeomorphism invariant measure. We close the section by explaining how the construction of the r-Fock measures extends to arbitrary $SU(N)$ gauge theory. We finally conclude in section IV with few comments and outlooks.

\section{The loop representations of gauge field theories}
We recall now the general idea of the "loop" construction of gauge invariant integrals on the spaces of gauge potentials (connections). Consider a manifold ${\cal M}$, a Lie group $G$ and its Lie algebra $\mathfrak{g}$. A $\mathfrak{g}$ valued $1$-form defined on ${\cal M}$ that is subject to the transformations 
\begin{align}\label{gauge0} 
A'= g^{-1}Ag + g^{-1}dg,\qquad g:{\cal M}\rightarrow G \ ,  
\end{align}
(for simplicity of the notation we think of $G$ as a subgroup of some $GL(N,\mathbb{C})$) is said to be a gauge potential while the transformations \eqref{gauge0} are referred to as gauge transformations. The space of the gauge potentials will be denoted by $\cal A$ and the space of the gauge transformations by $\cal G$. Considering gauge invariant functions on $\cal A$ amounts to dealing with the quotient space ${\cal A}/{\cal G}$.  The globally defined gauge potentials on ${\cal M}$ are related to connections defined on a trivial principal fiber bundle over ${\cal M}$ with the structure group $G$. The loop integral theory is available also for non-trivial principal fiber bundles (see \cite{Baez, AshLew}), however  trivial bundles  only feature in the current paper.    

Given a gauge potential $A$ and a curve\footnote{We assume that the manifold ${\cal M}$ is semi-analytic and that the curves we consider on ${\cal M}$ are piecewise analytic \cite{LQG3}.} $\gamma:[s_I,s_F]\rightarrow {\cal M}$, we define the corresponding parallel transport, or holonomy, $h_\gamma\in G$ as
\begin{align}
h_\gamma(A)  :=  \Pexp \left[ \int_{\gamma} A\right] \ ,
\end{align}
where $\Pexp$ stands for the path ordered exponential, that is
\begin{align}
h_\gamma(A) = I +\int_{\gamma} ds_1 A_a(\gamma(s_1)) \dot\gamma^a(s_1) +\int_{s_I}^{s_F} ds_1 \int_{s_I}^{s_1} ds_2 \ \dot\gamma^a(s_1) \dot\gamma^b(s_2) A_a(\gamma(s_1))A_b(\gamma(s_2)) + \dots 
\end{align}
The holonomy is insensitive to orientation preserving reparametrizations of $\gamma$, and, given a metric tensor $q$ on $M$,  without any loss of generality we can assume that the tangent vector $\dot \gamma(s) := d\gamma(s)/ds$ is normalized as $\dot\gamma^a(s)\dot\gamma^b(s)q_{ab} = 1$. Furthermore, given a curve $\gamma$ and the curve $\gamma^{-1}$ obtained by flipping the orientation, the corresponding parallel transport is
\begin{align}\label{hol1}
h_{\gamma^{-1}}(A)  = \left(h_\gamma\left(A\right)\right)^{-1}\ .
\end{align}
The holonomy satisfies the natural composition law with respect to the composition of curves, namely
\begin{align}\label{hol2}
h_{\gamma'\circ\gamma}(A)  = h_{\gamma'}(A)h_{\gamma}(A)\ .
\end{align}
Upon the gauge transformations \eqref{gauge0} the parallel transport transforms as follows
\begin{equation}\label{HolGauge} 
h_\gamma(A') = g^{-1}(\gamma(s_I)) h_\gamma(A)g(\gamma(s_F))\ . 
\end{equation}
To construct the loop representation \cite{LQG0, LQG1, LQG2, LQG3, LQG4} for a gauge theory with a compact gauge group $G$, the configuration space is taken to be the space of generalized connections $\bar{\cal A}$, which is an enlargement of the space ${\cal A}$ of smooth connections to include non-continuous connections that still assign well defined holonomies to curves in the spatial manifold. More precisely, every map $\gamma \mapsto h_\gamma(\bar{A})\in G$, which satisfies the properties \eqref{hol1} and \eqref{hol2} is considered a generalized connection $\bar{A}$, and $\bar{\cal A}$ is the space of such connections.
Then one takes the space $Cyl(\bar{\cal A})$ of cylindrical functions on $\bar{\cal A}$ to be the space of quantum states. An element $\Psi_\Gamma$ of $Cyl(\bar{\cal A})$ is a complex valued function $\psi\in C^0(G^n)$ which only depends on a finite set of holonomies, of which the curves (edges) form a graph $\Gamma$,
\begin{align}
\Psi_\Gamma[A]:= \psi(h_{e_1}(A),\dots,h_{e_n}(A))\ ,
\end{align}
where $\Gamma:=\cup_{i=1}^n e_i$. We say that the function $\Psi_\Gamma$ is a cylindrical function with respect to the graph $\Gamma$. A cylindrical function with respect to a given graph $\Gamma$, is also cylindrical with respect to any larger graph which contains $\Gamma$ as a sub-graph, and from \eqref{HolGauge} it follows that gauge transformations act on cylindrical functions by acting only at the vertices of the graphs. 

A linear map 
$\int : Cyl(\bar{\cal A}) \rightarrow \mathbb{C}$
is said to be an integral if
\begin{align}\label{positive}
\forall \Psi\in Cyl(\bar{\cal A}),\quad \int \overline{\Psi} = \overline{\int \Psi}\ , \ \int \overline{\Psi}\Psi \geq 0\ , \ \abs{\int \Psi } \leq C\norm{\Psi} \ ,
\end{align} 
with $C$ being a fixed positive real number independent of $\Psi$. Every integral $\int$ provides an integral $\int d\mu_\Gamma$ defined on $C^0(G^n)$ for every graph $\Gamma$, $n$ being the number of edges in $\Gamma$, and hence gives rise to a family of integrals $\{\int d\mu_\Gamma: \text{$\Gamma$ a graph in ${\cal M}$} \}$. This observation was used to define the first examples of faithful measures on $\bar{\cal A}$ \cite{ALmeasure, MarolfMourao}.  

While the action of gauge transformations extends naturally to $\bar{\cal A}$, the group $\cal G$ admits a natural extension to $\bar{\cal G} := G^M$, that is the space of all $G$ valued functions on $M$, not necessarily differentiable nor even continuous. 

When $G$ is a compact group, the structures defined above have quite strong properties. In particular, both spaces $\bar{\cal A}$ and $\bar{\cal G}$ are compact with respect to natural topologies, and the space $Cyl(\bar{\cal A})$ forms a dense subalgebra of the algebra  $C^0(\bar{\cal A})$ with respect to the sup-norm. 

Every integral defined on $Cyl(\bar{\cal A})$ defines a Borel measure on $\bar{\cal A}$, and gives rise to a family of Borel measures $\{(G^n,\mu_\Gamma): \text{$\Gamma$ a graph in ${\cal M}$} \}$. 
A natural diffeomorphism invariant measure $\mu^o$ is defined on  $\bar{\cal A}$ \cite{ALmeasure, MeasuresGT,MarolfMourao} such that every measure $\mu^o_\Gamma$ in the corresponding family of measures is given by
\begin{align}
\mu^o_\Gamma = \mu^H_G\otimes ... \otimes \mu^H_G\ ,
\end{align}
where $\mu^H_G$ is the normalised Haar measure on $G$.

The natural measure $\mu^o$ is used to define the kinematical Hilbert space ${\cal H}$ of loop quantum gravity as the space of square integrable functions with respect to the measure $\mu^o$:
\begin{align}
{\cal H}:= L^2(\bar{\cal A},\mu^o)\ .
\end{align}
It is useful to also introduce the space $Cyl^*(\bar{\cal A})$ which is defined as the algebraic dual of the space $Cyl(\bar{\cal A})$ of cylindrical functions. We then have
\begin{align}
Cyl(\bar{\cal A}) \subset {\cal H} \subset Cyl^*(\bar{\cal A}) \ .
\end{align}
The scalar product defined on $Cyl(\bar{\cal A})$ can be naturally transferred to the dual space $Cyl^*(\bar{\cal A})$.
The space $Cyl^*(\bar{\cal A})$ plays an important role in the context of the background independent theory as it allows the implementation of the spatial diffeomorphism constraint, and as we point out later, $Cyl^*(\bar{\cal A})$ is also the natural habitat of the r-Fock states when mapped to the loop theory.

When the local gauge invariance is imposed, that is the invariance with respect to the action of the gauge transformations in $\bar{\cal G}$, one obtains the Hilbert space ${\cal H}^G$ of $G$-gauge invariant states. The Hilbert space ${\cal H}^G$ admits an orthonormal basis, where each element is called a generalized spin network, or a $G$-colored network. A $G$-colored network is characterized by an embedded non-oriented graph\footnote{From here on, a graph actually consists of an equivalence class of embedded graphs where each graph can be obtained from another via a sequence of the following moves: splitting an edge, trivially connecting two edges, or changing the orientation of an edge.} $\Gamma$, a set of labels $\{j\}$ associated to the oriented edges of $\Gamma$ and corresponding to non trivial irreducible representations of the group $G$, and a set of $G$ gauge invariant tensors (intertwiners) $\{\iota\}$ associated to the vertices. To each vertex $v$ is associated a finite dimensional intertwiner space given by the tensor product of the representations assigned to the edges whose source is $v$ and the representations dual to those assigned to the edges whose target is $v$, and an intertwiner at $v$ is an invariant tensor in this space. It follows that ${\cal H}$ can be decomposed as a direct sum
\begin{align}\label{Decomp}
{\cal H}^G = \bigoplus_{\Gamma} {\cal H}_\Gamma^G = \bigoplus_{\Gamma, \{j\}} {\cal H}_{\Gamma, \{j\}}^G \ .
\end{align}
In the case $G=U(1)$, the non trivial irreducible representations can be labeled by integers, the intertwiners are trivial and the gauge invariance translates into the condition that, at each vertex of a graph, the sum of the integers labeling the edges vanishes. We denote the $U(1)$-colored networks ${\cal{N}}_{\Gamma,\vec{n}}$, where the $\Gamma$ stands for the graph and $\vec{n}$ the set of colors (integers) associated to the edges. Note that the decomposition \eqref{Decomp} can be generalized to the non gauge invariant spaces $Cyl(\bar{\cal A})$ and ${\cal H}$, and it amounts to introducing additional labelling of the vertices of the graphs by inequivalent irreducible representations of $G$.

Another way to recover the space of gauge invariant functions is to work from the start with gauge invariant quantities. 

Let $\mathcal{L}$ be the space of oriented, piecewise analytic loops on ${\cal M}$.
Thanks to the fact that the holonomy along a closed loop $\gamma$ depends on the gauge transformations at one point only, 
\begin{equation}\label{gauge1} 
h_\gamma(A') = g^{-1}(x_0) h_\gamma(A)g(x_0), \ \ \ x_0:=\gamma(s_0)=\gamma(s_1)\ . 
\end{equation}
it is easy to define gauge invariant functions. Indeed, given a loop $\gamma \in {\cal L}$,  we define the Wilson loop function
\begin{align}\label{Wilson}
W_\gamma(A) := {\rm Tr}(h_\gamma(A))\ . 
\end{align}
The Wilson loops generate a dense subalgebra in $Cyl(\bar{\cal A}/\bar{\cal G})$. Therefore, in order to determine a measure on $\bar{\cal A}/\bar{\cal G}$ it is enough to define an integral $\int$ on the algebra of polynomials of the Wilson loop functions. This amounts to constructing a linear map $\int$  which satisfies the conditions (\ref{positive}) for any arbitrary polynomial $\Psi$ of the Wilson loops. The resulting integral would extend to all the subalgebra of the gauge invariant cylindrical functions $Cyl(\bar{\cal A}/\bar{\cal G})$, hence it would define a measure on $\bar{\cal A}/\bar{\cal G}$. 

Note that given a measure $\mu$ on $\bar{\cal A}/\bar{\cal G}$, it is easy to construct a gauge invariant measure on $\bar{\cal A}$. For this purpose, one uses the Haar measure $\mu^H_{\bar{\cal G}}$ defined on the group ${\bar{\cal G}}$ (as every topological compact group admits a normalised Haar measure). The gauge invariant measure corresponding to $\mu$ measure on $\bar{\cal A}$ is then defined by the following integral
\begin{align}\label{GaugeInvMeasure}
\int \Psi := \int_{\bar{\cal A}/\bar{\cal G}} d\mu \int_{\bar{\cal G}} \mu^H_{\bar{\cal G}}\ \Psi\ . 
\end{align}

Our method to generalize the construction of r-Fock measures consists in embedding the Wilson loop functions in the space of operators defined on the Fock space, then using the Fock vacuum to define a new measure on $\bar{\cal A}/\bar{\cal G}$, lift it to a gauge invariant measure on $\bar{\cal A}$, and finally characterize it in terms of graphs and the corresponding measures.

\section{The \lowercase{r}-Fock representation for $U(1)$ gauge theory}

In this section, we briefly review the construction of the r-Fock representation for the Abelian $U(1)$ gauge theory, as developed in \cite{Var1}.

\subsection{The classical algebras}
Consider a $U(1)$ gauge theory on Minkowski spacetime. The phase space variables are the $U(1)$ connection 1-form $A_a$ and the conjugate electric field $E^b$ which satisfy the Poisson algebra
\begin{equation}\label{AEalg}
\{ A_{a}({x}), E^b ({y})\} = \frac{1}{q} \delta_a^b \delta^{(3)} ({x},{y})\ ,
\end{equation}
where $q$ is the charge parameter,  $\delta_a^b$ is the Kronecker delta, $\delta^{(3)} ({x},{y})$ is the Dirac delta distribution and $x,\ y$ denote coordinates on $\mathbbm{R}^3$.

Given $\gamma \in \mathcal{L}$, the holonomy of $A_a(x)$ along $\gamma$ is 
\begin{align}
 h_\gamma (A):= \exp \left[i \int_\gamma ds\ A_a(\gamma(s)) \dot{\gamma}^a(s) \right]\ ,
\end{align}
which can be equivalently defined as 
\begin{align}
h_\gamma(A)  =  \exp \left[ i\int_{R^3}d^3x\ X^a_\gamma({x}) A_a({x}) \right]\ ,
\end{align}
with the form factor function
\begin{align}
X^a_\gamma({x}) := \int_\gamma ds\ \delta^{(3)}(\gamma(s), {x}){\dot{\gamma}}^a(s) \ .
\end{align}
We then define the smeared electric field $E^a_r(x)$ as 
\begin{align}
E^a_r({x}) := \int d^3y f_r(x-y) E^a({\vec y}) \ ,
\label{eq:defer}
\end{align}
where $f_r({x})$ is a real valued Schwartz function\footnote{A further, and rather specific, restriction on the smearing function is obtained when analyzing a result derived later on in the present article. We refer the reader to the appendix \ref{AppendixA} for more details.} which approximates the Dirac delta function for small $r$:
\begin{align}
 \forall x,y\in \mathbbm{R}^3,\ \lim_{r\rightarrow 0} f_r(x-y) = \delta^{(3)}(x,y)\ .
\end{align}
It follows that the holonomies and the smeared electric field satisfy
\begin{align}\label{HolSEalg}
 \{ h_{\gamma},h_{\alpha}\} = \{ E^a_r(x), E^b_r({\vec y})\} = 0 \ , \ 
 \{ h_{\gamma}, E^a_r(x)\} = \frac{i}{q} X^a_{\gamma,r}(x) h_{\gamma}\ .
\end{align}

In the Fock representation of the Poisson algebra generated by \eqref{AEalg}, the holonomy operator is not a well defined. Neither is the connection in the loop representation of the holonomy algebra. Therefore, in order to make a contact between the two representations, M.\ Varadarajan introduced in \cite{Var1} the so called r-Fock representation of the holonomy algebra, obtained from the Fock representation of the \emph{smeared holonomy} algebra which is defined as follows.

Let us introduce the smearing form factor
\begin{align}
 X^a_{\gamma,r}({x}) := \int_{R^3} d^3y\ f_r(x - y) X^a_{\gamma}({y}) = \int_{\gamma} ds f_r(x - \gamma(s)){\dot{\gamma}}^a(s)\ ,
\end{align}
and the smeared connection $A_a^r$,
\begin{align}
 A_a^r (x) = \int_{R^3} d^3y f_r(x-y) A_a(y)\ .
\end{align}
A smeared holonomy associated to a loop $\gamma \in \mathcal{L}$ is defined as
\begin{align}
 h_\gamma^r(A) := \exp \left[ i \int_\gamma ds\ \dot{\gamma}^a(s)  A_a^r(\gamma(s)) \right] = \exp \left[ i \int_{R^3} d^3x\ X^a_{\gamma,r}({x})  A_a(x) \right]\ .
\end{align}
Along with the electric field $E^a({x})$, they satisfy
\begin{align}\label{SHolEalg}
\{ h_\gamma^r,h_\alpha^r\} = \{ E^a({x}), E^b({y})\} = 0\ ,\ \{ h_\gamma^r, E^a({x})\} = \frac{i}{q} X^a_{\gamma,r}({x}) h_{\gamma}^r\ .
\end{align}
It was shown in \cite{Var1} that the Fock representation of the algebra $\cal{A}$ generated by \eqref{AEalg} is a representation of the Poisson bracket algebra ${\cal{HA}}_r$ generated by \eqref{SHolEalg}. Additionally, unlike the standard holonomies, the quantized smeared holonomies ${\hat h}_{\gamma}^r$ are unitary operators in the standard Fock representation.

Furthermore, it was also shown that the Poisson algebra ${\cal{HA}}_r$ generated by $(h_{\gamma}^r(A) , E^a(x))$ and the Poisson algebra $\cal{HA}$ generated by $(h_{\gamma}(A) , E^a_r(x))$ are isomorphic. This means that any representation of the algebra ${\cal{HA}}_r$ is a representation of the algebra $\cal{HA}$. In particular the Fock representation of the algebra ${\cal{HA}}_r$ is a representation of the algebra $\cal{HA}$. This representation is what is called the r-Fock representation of $\cal{HA}$.

\subsection{The r-Fock representation}

Since the standard Fock representation can be reconstructed from the expectation values of the algebra operators in the vacuum state, the strategy is to define the r-Fock representation via the vacuum expectation values in the Fock representation. This goes as follows.

Using the standard Fock quantization of the connection and electric field
\begin{align}
 \hat A_a (x) &:= \frac{1}{(2 \pi)^{3/2}} \int \frac{d^3k}{q\sqrt{2|k|}} \left( e^{-i \vec{k}.\vec{x}}\ c_a^\dag (k) + e^{i \vec{k}.\vec{x}}\ c_a (k) \right)\\
  \\
 \hat E^a(x) &:= \frac{i}{(2 \pi)^{3/2}} \int d^3k \sqrt{\frac{|k|}{2}} \left( e^{ \vec{k}.\vec{x}}\ c_a^\dag (k) - e^{i \vec{k}.\vec{x}}\ c_a (k) \right)\ ,
\end{align}
where $c_a$ and $c_a^\dag$ are the annihilation and creation operators respectively, satisfying
\begin{align}\label{CanCom}
 [c_a(k) , c_b^\dag(l)] = \delta_{ab}\ \delta^{(3)}(k,l)\ ,
\end{align}
the smeared connection operator is
\begin{align}
 \hat A_a^r (x) &:= \frac{1}{(2 \pi)^{3/2}} \int \frac{d^3k}{q\sqrt{2|k|}} \int d^3y f_r(x-y) \left( e^{-i \vec{k}.\vec{y}}\ c_a^\dag (k) + e^{i \vec{k}.\vec{y}}\ c_a (k) \right)\\
  \\
 &= \int \frac{d^3k}{q\sqrt{2|k|}} \left( e^{-i \vec{k}.\vec{x}}\ \tilde{f}_r(k)\ c_a^\dag (k) + e^{i \vec{k}.\vec{x}}\ \overline{\tilde{f}_r(k)}\ c_a (k) \right)\ ,
\end{align} 
where $\tilde{f}_r(k)$ denotes the Fourier transform of $f_r(x)$.
Hence, the expression of the smeared holonomy operator is
\begin{align}\label{S.Hol.Op.U1}
 \hat{h}_\gamma^r(A) :=& \exp \left[ i \int_\gamma ds\ \dot{\gamma}^a(s) \hat{A}_a^r(\gamma(s)) \right] \\
 =& \exp \left[ i \int_\gamma ds\ \dot{\gamma}^a(s) \int \frac{d^3k}{q\sqrt{2|k|}} \left( e^{-i \vec{k}.\vec{\gamma}(s)}\ \tilde{f}_r(k)\ c_a^\dag (k) + e^{i \vec{k}.\vec{\gamma}(s)}\ \overline{\tilde{f}_r(k)}\ c_a (k) \right) \right]\\
 =& \exp \left[ i \int \frac{d^3k}{q\sqrt{2|k|}} \left( \tilde{X}^a_{\gamma,r}(k)\ c_a^\dag (k) + \overline{\tilde{X}^a_{\gamma,r}(k)}\ c_a (k) \right) \right]\ ,
\end{align}
where $X^a_{\gamma,r}(k)$ denotes the Fourier transform of $X^a_{\gamma,r}(x)$.

The r-Fock representation of $\cal{HA}$ is then specified via the expectation values of the smeared holonomy and the electric field operators in the standard Fock vacuum, denoted $\ket{0}$, namely
\begin{align}\label{HolExpValU1}
& \bra{0}{\hat h}_\gamma^r\ket{0} = \exp (-\int \frac{d^3k}{4q^2 |k|} |X^a_{\gamma,r}(k)|^2) =: \bra{0_r}{\hat h}_{\gamma}\ket{0_r} \\
\ \\
& \bra{0}{\hat h}_\alpha^r{\hat E}^a(x){\hat h}_\beta^r\ket{0} = \frac{X^a_{\beta,r}(x)-X^a_{\alpha,r}(x)}{2q} \exp (-\int \frac{d^3k}{4q^2 |k|} |X^a_{{\alpha\circ \beta},r}(k)|^2)
=: \bra{0_r}{\hat h}_{\alpha}{\hat E}^a_r(x){\hat h}_{\beta}\ket{0_r}\ ,
\end{align}
for arbitrary loops $\gamma, \ \alpha,\ \beta$ in $\mathcal{L}$.

Through equation \eqref{HolExpValU1}, the holonomy operators ${\hat h}_{\gamma}$ are introduced as well defined operators in the r-Fock representation. This fact, as shown in the following section, allows to explicitly relate the r-Fock representation and the loop representation as two inequivalent representations of the algebra ${\cal{HA}}$.

\subsection{Relating the r-Fock measure and the natural measure for the U(1) gauge theory}

Using the results above, we can explicitly define the r-Fock measure on the space of cylindrical functions $Cyl\big(\bar{\cal A}/\bar{\cal G}\big)$ in the case of the $U(1)$ gauge theory. This goes as follows \cite{Var1, Var2}.

Given a $U(1)$-colored network state ${\cal N}_{\Gamma, \vec{n}}$, with graph $\Gamma$ given by a set of oriented edges $E(\Gamma)$, where to each edge $e_I$ is associated a representation label, that is an integer $n_I$, the corresponding cylindrical function provides an operator $\hat \Psi$ on the Fock space of the $U(1)$ Fock representation defined via smeared holonomy operators as
\begin{align}
 \hat \Psi_{\Gamma, \vec{n}} :=& \prod_{I\in E(\Gamma)}\hat{h}_{e_I}^r \ .
\end{align}
From the definition of the smeared holonomy operator \eqref{S.Hol.Op.U1}, one obtains an expression for $\hat \Psi$ in terms of the canonical operators, namely
\begin{align}
 \hat \Psi_{\Gamma, \vec{n}} =& \prod_{I\in E(\Gamma)} \exp \left[ i\ n_I \int_{e_I} ds\ \dot{e}_I^a(s) \hat{A}_a^r(e(s)) \right] \\
 =& \exp \left[ i \sum_I n_I \int_{e_I} ds\ \dot{e}_I^a(s) \hat{A}_a^r(e(s)) \right] \\
  =& \exp \left[ i \sum_I n_I \int_{e_I} ds\ \dot{e}_I^a(s) \int \frac{d^3k}{q\sqrt{2|k|}} \left( e^{-i \vec{k}.\vec{e}(s)}\ \tilde{f}_r(k)\ c_a^\dag (k) + e^{i \vec{k}.\vec{e}(s)}\ \overline{\tilde{f}_r(k)}\ c_a (k) \right) \right]\\
 =& \exp \left[ i \int \frac{d^3k}{q\sqrt{2|k|}} \left(\sum_I n_I \tilde{X}^a_{e_I,r}(k)\right) c_a^\dag (k) \right] \exp \left[ i \int \frac{d^3k}{q\sqrt{2|k|}} \left(\sum_I n_I \overline{\tilde{X}^a_{e_I,r}(k)}\right) c_a (k) \right] \times \\
 & \exp \left[ -\frac{1}{4q^2} \sum_{IJ} n_I n_J \int \frac{d^3k}{|k|}\ \tilde{X}^a_{e_I,r}(k)\ \overline{\tilde{X}^a_{e_J,r}(k)} \right]\ .
\end{align}

The r-Fock measure $\mu_{U(1)}^r$ on the space of cylindrical function as well as the r-Fock vacuum $\ket{0_r}$ are then defined as
\begin{align}
  &\forall {\cal N}_{\Gamma, \vec{n}} \in Cyl\big(\bar{\cal A}/\bar{\cal G}\big),\\
  &\int_{\bar{\cal A}/\bar{\cal G}} d\mu_{U(1)}^r\ {\cal N}_{\Gamma, \vec{n}}(A) := \bra{0_r}{\hat {\cal N}_{\Gamma, \vec{n}}}\ket{0_r} := \bra{0} \hat \Psi_{\Gamma, \vec{n}} \ket{0} = \exp \left[ -\frac{1}{4q^2} \sum_{I,J} n_I n_J \int \frac{d^3k}{|k|}\ \tilde{X}^a_{e_I,r}(k)\ \overline{\tilde{X}^a_{e_J,r}(k)} \right]\ .
\end{align}
Thanks to the decomposition \eqref{Decomp} and the fact that the natural measure $\mu_{U(1)}^o$ satisfies
\begin{align}
 \forall {\cal N}_{\Gamma, \vec{n}} \in Cyl\big(\bar{\cal A}/\bar{\cal G}\big),\ \int_{\bar{\cal A}/\bar{\cal G}} d\mu_{U(1)}^o\ {\cal N}_{\Gamma, \vec{n}}(A) =\left\{
 \begin{array}{l}
 1,\ \text{if}\ \Gamma = \tilde 0\\
 0,\ \text{if}\ \Gamma \neq \tilde 0
 \end{array}
 \right. \ ,
\end{align}
where $\tilde 0$ stands for the trivial graph equivalence class characterized by no graph, one can relate the r-Fock measure $\mu_{U(1)}^r$ to the measure $\mu^o$ on $Cyl\big(\bar{\cal A}/\bar{\cal G}\big)$ and one gets
\begin{align}
 d\mu_{U(1)}^r = \left(\sum_{\Gamma, \vec{n}} \exp \left[ -\frac{1}{4q^2} \sum_{I,J} n_I n_J \int \frac{d^3k}{|k|}\ \tilde{X}^a_{e_I,r}(k)\ \overline{\tilde{X}^a_{e_J,r}(k)} \right]\overline{{\cal{N}}_{\Gamma,\vec{n}}} \right) d\mu_{U(1)}^o \ .
\end{align}

We conclude this section about the $U(1)$ r-Fock representation by pointing out that the r-Fock states can be identified as states in the loop representation, in particular the r-Fock vacuum $\ket{0_r}$. For the r-Fock vacuum $\ket{0_r}$, one can use the Poincare invariance of the state \cite{Var2}, i.e.\ the fact that $\ket{0_r}$ is annihilated by the operators $c_a(k)$. This leads to the identification of a state ${\cal V}_F^r$ corresponding to $\ket{0_r}$ which does not belong to the Hilbert space ${\cal H}$ but rather to the space $Cyl^*$. It is a distributional state which acts on the states in ${\cal H}$ and has the following expression 
\begin{align}
 {\cal V}_F^r &= \sum_{\Gamma, \vec{n}} \exp \left[ -\frac{1}{q^2} \sum_{I,J \in E(\Gamma)} n_I n_J \int \frac{d^3k}{2|k|}\ \tilde{X}^a_{e_I,r}(k)\ \overline{\tilde{X}^a_{e_J,r}(k)} \right]\bra{{\cal{N}}_{\Gamma,\vec{n}}} \ .
\end{align}

In the next section, we present a method to generalize the construction of the r-Fock measure to the case of the non Abelian $SU(N)$ gauge theory.

\section{The \lowercase{r}-Fock measure for a non-Abelian $SU(N)$ gauge theory}

In the case of $U(1)$ gauge theory discussed above, as well as any Abelian gauge theory, the construction of the r-Fock measure is based on the identification
\begin{align}\label{IdentU1}
 \forall \gamma \in {\cal{L}}, \qquad \int d\mu_{U(1)}^r\ h_\gamma(A) := \bra{0}\hat h_\gamma^{r} \ket{0} \ ,
\end{align}
which is sufficient to define the measure and its properties. In particular, it implies that for every $U(1)$ cylindrical function $\Psi \in Cyl\big(\bar{\cal A}/\bar{\cal G}\big)$ such that $\Psi(A) := \psi(h_{\gamma_1}(A),\dots,h_{\gamma_K}(A))$, we have
\begin{align}\label{GenIdentU1}
 \int_{\bar{\cal A}/\bar{\cal G}} d\mu_{U(1)}^r\ \Psi(A) := \bra{0}\psi(\hat h_{\gamma_1}^{r}, \dots, \hat h_{\gamma_K}^{r}) \ket{0} \ .
\end{align}
This fact can be understood as a consequence of the Abelian nature of the $U(1)$ group, but because we know that in the case of $U(1)$ we have $W_\gamma(A) = h_\gamma(A)$, this result can also be understood as a consequence of Mandelstam identities \cite{GamTri, LQG0} for $U(1)$, which namely imply that every $U(1)$ cylindrical function can be expressed as a linear combination of Wilson loops.

Our aim now is to define an $r$-Fock measure $\mu^r$ on the space of cylindrical functions for a non-Abelian $SU(N)$ gauge theory on Minkowski spacetime by generalizing the identification \eqref{IdentU1}. Mandelstam identities for $SU(N)$ imply that the natural generalization of \eqref{IdentU1} takes the form
\begin{align}\label{IdentSUN}
 \forall \{\gamma_1,\dots,\gamma_{N-1}\} \in {\cal{L}}^{N-1}, \quad \int d\mu_{SU(N)}^r\ W_{\gamma_1}^J(A) \dots W_{\gamma_{N-1}}^J(A) := \bra{0}\hat W_{\gamma_1}^{r,J} \dots \hat W_{\gamma_{N-1}}^{r,J} \ket{0}\ ,
\end{align}
where $W_{\gamma_i}^J$ is the Wilson loop in the irreducible representation $J$ associated to the loop $\gamma_i$, $\ket{0}$ is the Fock vacuum in the $SU(N)$ gauge theory, and $\hat W_{\gamma_i}^{r}$ is an operator associated to the same loop $\gamma_i$ and acting on the Fock space of the $SU(N)$-gauge theory given by
\begin{align}
 \hat W_\gamma^{r,J} := \Tr \left[ \hat h_\gamma^{r,J} \right] \ ,
\end{align}
with $\hat h_\gamma^{r,J}$ being the smeared holonomy operator which we define later. We call the operator $\hat W_\gamma^{r,J}$ the r-Wilson loop operator in the $J$ representation. Note, however, that in order to define the r-Fock measure, it is sufficient to establish the identification \eqref{IdentSUN} in the fundamental representation of the gauge group $SU(N)$, as the results for the other representations can be in principle derived via recoupling theory. 

Computing the expectation value of a product of r-Wilson loop operators in the Fock vacuum for a non Abelian $SU(N)$ gauge theory is not as straightforward nor as explicit as in the Abelian case. Therefore, we first focus on the construction for the particular case of $SU(2)$, the simplest non Abelian $SU(N)$ group, then we present how to generalize the calculations and the results to arbitrary $SU(N)$ groups.

\subsection{Fock vacuum expectation value of the $SU(2)$ r-Wilson loop operator}

Using a similar notation as in the $U(1)$ case, the phase space variables of the $SU(2)$ gauge theory are a $\mathfrak{su}(2)$ Lie algebra valued connection $A_a^i$ and the conjugate field $E_j^b$ satisfying
\begin{equation}
\{ A_{a}^i({x}), E_j^b ({y})\} = \frac{1}{q} \delta_j^i \delta_a^b \delta^{(3)} ({x},{y})\ .
\end{equation}
Smearing the connection $A$ gives the $\mathfrak{su}(2)$ Lie algebra valued 1-form\footnote{Thanks to the fact that all $SU(N)$ bundles over a 3-manifold are trivial, we fix, once and for all, a global trivialization which allows to regard all smooth $\mathfrak{su}(N)$ Lie algebra valued 1-forms on $\mathbbm{R}^3$ as $SU(N)$-connections on a trivial bundle over the spatial manifold.} $A^r$:
\begin{align}\label{Ar}
A_a^{r,i} (x) = \int_{R^3} d^3y f_r(x-y) A_a^i(y)\ ,
\end{align}
and a smeared holonomy associated to a loop $\gamma \in \mathcal{L}$ in an arbitrary representation labeled by the spin $J$ is then defined as
\begin{align}\label{SU(2)_smeared_hol}
h_\gamma^{r,J}(A) :=& \Pexp \left[ \int_\gamma ds\ \dot{\gamma}^a(s) A_a^{r,i}(\gamma(s)) \tau_i^{(J)}  \right] \ ,
\end{align}
where $\tau_i^{(J)}$ represents the three $SU(2)$ generators in the irreducible representation $J$. We denote by $W_\gamma^{r,J}$ the trace of the smeared holonomy $h_\gamma^{r,J}$.

Note that unlike the Abelian case, local gauge transformations acting on the connection $A$ do not correspond to local gauge transformations acting on the connection $A^r$. Therefore, the smeared holonomies transform covariantly under local gauge transformation acting on $A^r$, but do not transform covariantly under local gauge transformation acting on $A$.

Together with the electric field $E_j^b$, the smeared holonomies satisfy
\begin{align}
\{ h_\gamma^{r,J},h_\alpha^{r,J}\} = \{ E_i^a({x}), E_j^b({y})\} = 0\ ,\ \{ h_\gamma^{r,J}, E_i^a({x})\} = \frac{1}{q} X^a_{\gamma,r}({x})  h_{\gamma(1,x)}^{r,J} \tau_i^{(J)} h_{\gamma(x,0)}^{r,J} \ .
\end{align}
In the standard Fock representation of an $SU(2)$ gauge theory on Minkowski spacetime, the smeared connection operator is given by
\begin{align}\label{Aop}
 \hat A_a^{r,i} (x) :=& \frac{1}{(2 \pi)^{3/2}} \int \frac{d^3k}{q\sqrt{2|k|}} \int d^3y f_r(x-y) \left( e^{-i \vec{k}.\vec{y}}\ c_a^{i \dag} (k) + e^{i \vec{k}.\vec{y}}\ c_a^i (k) \right)\\
 =& \int \frac{d^3k}{q\sqrt{2|k|}} \left( e^{-i \vec{k}.\vec{x}}\ \tilde{f}_r(k)\ c_a^{i \dag} (k) + e^{i \vec{k}.\vec{x}}\ \overline{\tilde{f}_r(k)}\ c_a^i (k) \right)\\
 =& \int \frac{d^3k}{q\sqrt{2|k|}} e^{-i \vec{k}.\vec{x}}\ \tilde{f}_r(k) \left( c_a^{i \dag} (k) + c_a^i (-k) \right) \ ,
\end{align} 
with
\begin{align}\label{CanCom2}
 [c_a^i(k) , c_b^{j \dag}(l)] = \delta_{ab}\delta^{ij} \delta^{(3)}(k,l) \ .
\end{align}

Now consider the $SU(2)$ smeared holonomy operator defined as
\begin{align} \label{SU2-r-hol}
 \hat{h}_\gamma^{r,J} :=& \Pexp \left[ \int_\gamma ds\ \dot{\gamma}^a(s) \hat{A}_a^{r,i}(\gamma(s)) \tau_i^{(J)}  \right] \\ 
 =& \Pexp \left[ \int_\gamma ds \int \frac{d^3k}{q\sqrt{2|k|}} \tau_i^{(J)}  \left( \tilde{X}^a_{\gamma,r}(s,k) \ c_a^{i \dag} (k) + \overline{\tilde{X}^a_{\gamma,r}(s,k)} \ c_a^i (k) \right) \right]\\
 =& \Pexp \left[ \int_\gamma ds \int \frac{d^3k}{q\sqrt{2|k|}} \tau_i^{(J)} \tilde{X}^a_{\gamma,r}(s,k) \left( c_a^{i \dag} (k) + c_a^i (-k) \right) \right]\ ,
\end{align}
which consists of a matrix of operators, each acting in the Fock space, where
\begin{align}\label{X.func}
 \tilde{X}^a_{\gamma,r}(s,k):= \dot{\gamma}^a(s) e^{-i \vec{k}.\vec{\gamma}(s)}\ \tilde{f}_r(k)\ .
\end{align}
We then define the r-Wilson loop operator acting in the Fock space as
\begin{align}\label{r-Wilson}
 \hat{W}_\gamma^{r,J} :=& \Tr \left[ \hat{h}_\gamma^{r,J} \right]
 = \Tr \left[ \Pexp \left[ \int_\gamma ds \int \frac{d^3k}{q\sqrt{2|k|}} \tau_i^{(J)}  \tilde{X}^a_{\gamma,r}(s,k) \left( c_a^{i \dag} (k) + c_a^i (-k) \right) \right] \right]\ .
\end{align}
Since the operators $\hat A_a^{r,i} (x)$ in \eqref{Aop} commute with each other, it follows that the r-Wilson loop operators also commute with each other, namely
\begin{align}
 \forall \gamma_1,\gamma_2 \in \mathcal{L},\ \forall J_1, J_2, \qquad \left[ \hat W_{\gamma_1}^{r,J_1} , \hat W_{\gamma_2}^{r,J_2} \right] = 0\ .
\end{align}
As mentioned earlier, establishing the identification \eqref{IdentSUN} in the fundamental representation, which in the $SU(2)$ case corresponds to $J=1/2$, is sufficient to define the measure. Furthermore, in case of $SU(2)$, the operator on the right-hand side of the equation in \eqref{IdentSUN} consists of a single r-Wilson loop operator.
Hence, the goal is to establish the general expression of the expectation value in the Fock vacuum of the r-Wilson loop operator defined in \eqref{r-Wilson} in representation $1/2$. 

For the explicit calculations of the expectation value of the r-Wilson loop in the standard Fock vacuum, our strategy consists of expressing the path ordered exponential in \eqref{r-Wilson} as its defining series expansion, and by using the linearity of the trace, we compute the trace of each term in the expansion. Then we compute the expectation value in the Fock vacuum of each term in the expansion, and eventually perform the summation of the series in order to obtain the final result.

We start with the expansion of the r-Wilson loop operator
{\small
\begin{align}\label{ExpansionWr}
 \hat{W}_\gamma^{r,1/2}
 &= \Tr \left[ \Pexp \left[ \int_\gamma ds \int \frac{d^3k}{q\sqrt{2|k|}} \tau_i  \tilde{X}^a_{\gamma,r}(s,k) \left( c_a^{i \dag} (k) + c_a^i (-k) \right) \right] \right]\\ 
 &= \Tr \left[ \sum \limits_{n=0}^\infty \int_0^1 ds_1\ \dots \int_0^{s_{n-1}} ds_n \prod \limits_{m=1}^n \int \frac{d^3k_m}{q\sqrt{2|k_m|}} \tau_{i_m} \tilde{X}^{a_m}_{\gamma,r}(s_m,k_m) \left( c_{a_m}^{i_m \dag} (k_m) + c_{a_m}^{i_m} (-k_m) \right) \right] \\ 
 &= \sum \limits_{n=0}^\infty \Tr \left[ \prod \limits_{m=1}^n \tau_{i_m} \right] \Pint{\gamma} ds_1 \dots ds_n \prod \limits_{m=1}^n \int \frac{d^3k_m}{q\sqrt{2|k_m|}} \tilde{X}^{a_m}_{\gamma,r}(s_m,k_m) \left( c_{a_m}^{i_m \dag} (k_m) + c_{a_m}^{i_m} (-k_m) \right)\ .
\end{align}
}
It follows that the expectation of the r-Wilson loop in the Fock vacuum gives
{\small
\begin{align}\label{ExpWr}
\hspace{-0,3cm}
 \mean{\hat{W}_\gamma^{r,1/2}}
 &= \sum \limits_{n=0}^\infty \Tr \left[ \prod \limits_{m=1}^n \tau_{i_m} \right] \Pint{\gamma} ds_1 \dots ds_n  \mean{ \prod \limits_{m=1}^n \int \frac{d^3k_m}{q\sqrt{2|k_m|}} \tilde{X}^{a_m}_{\gamma,r}(s_m,k_m) \left( c_{a_m}^{i_m \dag} (k_m) + c_{a_m}^{i_m} (-k_m) \right) } \\ 
 &= \sum \limits_{n=0}^\infty \Tr \left[ \prod \limits_{m=1}^{2n} \tau_{i_m} \right] \Pint{\gamma} ds_1 \dots ds_{2n} \left(\prod \limits_{m=1}^{2n} \int \frac{d^3k_m}{q\sqrt{2|k_m|}} \tilde{X}^{a_m}_{\gamma,r}(s_m,k_m) \right) \mean{ \prod \limits_{m=1}^{2n} \left( c_{a_m}^{i_m \dag} (k_m) + c_{a_m}^{i_m} (-k_m) \right)}
\end{align}
}
where in the second line we used Wick's theorem \cite{Wick} to eliminate from the sum the terms which contain the expectation value of an odd product of ladder operators, as it vanishes, leaving a sum over even terms only.
Now there are two general expressions to be evaluated separately: the first is the expectation value of the even product of ladder operators, the second is the trace of an even product of $\tau_i$ generators in representation $1/2$.

On one hand, the expectation value of the product of ladder operators can be computed using Wick's theorem and we get
\begin{align}\label{w2n}
 & \mean{ \prod \limits_{m=1}^{2n} \left( c_{a_m}^{i_m \dag} (k_m) + c_{a_m}^{i_m} (-k_m) \right)}\\ 
 & = \frac{1}{2^n n!}\sum_\sigma \scalemath{0.87}{ \delta_{a_{\sigma(1)}a_{\sigma(2)}} \dots \delta_{a_{\sigma(2n-1)}a_{\sigma(2n)}}
 \delta^{i_{\sigma(1)}i_{\sigma(2)}} \dots \delta^{i_{\sigma(2n-1)}i_{\sigma(2n)}}
 \delta^{(3)}(k_{\sigma(1)},-k_{\sigma(2)}) \dots \delta^{(3)}(k_{\sigma(2n-1)},-k_{\sigma(2n)})} ,
\end{align}
where the sum is over the permutations on a set of $2n$ elements.
On the other hand, given the $SU(2)$ generators $\tau_i$ in representation $1/2$ which satisfy
\begin{align}\label{taus}
\left[ \tau_i ,\tau_j \right]= \epsilon_{ijk} \tau_k\ ,\ \Tr[\tau_i \tau_j] = -\frac{1}{2} \delta_{ij} \ ,
\end{align}
where $\epsilon_{ijk}$ is the anti-symmetric Levi-Cevita tensor, the trace of the product of an even number of $\tau_i$ gives
\begin{align}
 \Tr \left[ \prod \limits_{m=1}^{2n} \tau_{i_m} \right] = \frac{(-1)^n}{2^{2n-1}} \sum_{\tilde \sigma} \text{sgn}(\tilde \sigma)\, \delta_{i_{\tilde \sigma(1)}i_{\tilde \sigma(2)}} \dots \delta_{i_{\tilde \sigma(2n-1)}i_{\tilde \sigma(2n)}}\ ,
\end{align}
where the sum is over the permutations $\tilde \sigma$ on a set of $2n$ elements, which map the ordered set $\{1,\dots,2n\}$ to an ordered set $\{\tilde \sigma(1),\dots, \tilde \sigma(2n)\}$ satisfying $\tilde \sigma(2s+1)<\tilde \sigma(2s+2)$ and $\tilde \sigma(2s+1)<\tilde \sigma(2s+3)$ for all valid integer values of $s$ (implying that $\tilde \sigma(1)=1$). For a set of $2n$ elements, there are $(2n-1)!!$ such permutations.

We then introduce the coefficients
\begin{align}\label{Y2n}
 \Upsilon_{\sigma^{(2n)}}^{(1/2)} := \delta^{i_{\sigma(1)}i_{\sigma(2)}} \dots \delta^{i_{\sigma(2n-1)}i_{\sigma(2n)}} \Tr \left[ \prod \limits_{m=1}^{2n} \tau_{i_m} \right]\ ,
\end{align}
which depend on the permutation $\sigma$ of a set of $2n$ elements, and the superscript $(1/2)$ is to recall the representation of the $\tau_i$ generators. The coefficients $\Upsilon_{\sigma^{(2n)}}^{(1/2)}$ are bounded in absolute value 
\begin{align}
\abs{\Upsilon_{\sigma^{(2n)}}^{(1/2)}}\leq 3^n/2^{2n-1}\ , 
\end{align}
and they satisfy
\begin{align}
 \sum_{\sigma} \Upsilon_{\sigma^{(2n)}}^{(1/2)} &= \Tr \left[ \prod \limits_{m=1}^{2n} \tau_{i_m} \right] \sum_{\sigma} \delta^{i_{\sigma(1)}i_{\sigma(2)}} \dots \delta^{i_{\sigma(2n-1)}i_{\sigma(2n)}} 
 = \frac{(-1)^n}{2^{2n-1}} (2n+1)!!\ .
\end{align}

Plugging \eqref{Y2n} and \eqref{w2n} into \eqref{ExpWr} gives
\begin{align} \label{FExpWr}
 \mean{\hat{W}_\gamma^{r,1/2}}
 &= \sum \limits_{n=0}^\infty \frac{1}{2^n n!}\Pint{\gamma} ds_1 \dots ds_{2n} \sum_\sigma \Upsilon_{\sigma^{(2n)}}^{(1/2)} \left(\prod \limits_{m=1}^{n} \int \frac{d^3k}{2 q^2 |k|} \tilde{X}^{a}_{\gamma,r}(s_{\sigma(2m-1)},k) \tilde{X}^{a}_{\gamma,r}(s_{\sigma(2m)},-k) \right)\ .
\end{align}
Because of the presence of the coefficients $\Upsilon_{\sigma^{(2n)}}^{(1/2)}$ for which we do not have a simplified expression, we cannot perform the sum in \eqref{FExpWr} and reduce it to a closed form. Therefore the expression in \eqref{FExpWr} is our final expression for the vacuum expectation value of a r-Wilson loop operator. As shown in the appendix \ref{AppendixA}, the sum in \eqref{FExpWr} is convergent and the expectation value $\mean{\hat{W}_\gamma^{r,1/2}}$ is always finite.

Thanks to the presence of the trace in the definition of the r-Wilson loop operator, the final result in \eqref{FExpWr} does not depend on choice of the $SU(2)$ generators $\tau_i$ as long as they satisfy (\ref{taus}). The gauge invariance requires an explanation: as we pointed out above, the smeared connections and their holonomies are not gauge covariant in any understood sense and the corresponding Wilson loop functions are not gauge invariant.  Nonetheless, all the expectation values $\mean{\hat{W}_\gamma^{r,1/2}}$ are manifestly gauge invariant.

\subsection{The Fock positive linear functional and the $SU(2)$ r-Fock measure}

The Fock quantization of ${\cal{HA}}_r$ can be achieved through the GNS construction based on the positive linear functional $\Phi_F$ induced by \eqref{FExpWr}:
\begin{align} \label{FPLF}
 \Phi_F \left[\sum \limits_{i=1}^{M}a_i W_{\gamma_i}^{r,1/2} \right] := \sum \limits_{i=1}^{M}a_i \mean{\hat{W}_{\gamma_i}^{r,1/2}}\ ,
\end{align}
for any finite set of $M$ arbitrary loops $\gamma_i \in {\cal L}$ and $M$ complex numbers $a_i$.
One can show that the linear functional $\Phi_F$ is positive on ${\cal{HA}}_r$:
\begin{align}
 \Phi_F \left[ \overline{\left(\sum \limits_{i=1}^{M}a_i W_{\gamma_i}^{r,1/2} \right)} \left(\sum \limits_{i=1}^{M}a_i W_{\gamma_i}^{r,1/2} \right) \right] &= \Phi_F \left[ \sum \limits_{i,j}^{M}\bar{a}_j a_i W_{\gamma_j}^{r,1/2} W_{\gamma_i}^{r,1/2} \right] \\ 
 &= \Phi_F \left[ \sum \limits_{i,j}^{M}\bar{a}_j a_i \left(W_{\gamma_j \circ \gamma_i}^{r,1/2} + W_{\gamma_j\circ \gamma_i^{-1}}^{r,1/2} \right) \right] \\
 &= \sum \limits_{i,j}^{M}\bar{a}_j a_i \left( \mean{\hat W_{\gamma_j \circ \gamma_i}^{r,1/2}} + \mean{\hat W_{\gamma_j\circ \gamma_i^{-1}}^{r,1/2}} \right)  \\
 &=  \sum \limits_{i,j}^{M}\bar{a}_j a_i \mean{\hat W_{\gamma_j}^{r,1/2} \hat W_{\gamma_i}^{r,1/2}}  \\
 &= \mean{ \left(\sum \limits_{j=1}^{M}a_j \hat W_{\gamma_j}^{r,1/2} \right)^\dag \left(\sum \limits_{i=1}^{M}a_i \hat W_{\gamma_i}^{r,1/2} \right) } \geq 0\ ,
\end{align}
where in the second and the forth lines we used one of the Mandelstam identities for the smeared Wilson loops and the r-Wilson loop operators respectively, and in the last line we used the self-adjointness of the r-Wilson loop operators. Those properties are discussed in the appendixes \ref{AppendixB} and \ref{AppendixC}.

Using the positive linear functional $\Phi_F$ on ${\cal{HA}}_r$, we define the $SU(2)$ r-Fock measure on the space $\bar{\cal A}/\bar{\cal G}$ as
\begin{align}\label{rFockMeasure}
 \int_{\bar{\cal A}/\bar{\cal G}} d\mu_{SU(2)}^r \sum \limits_{i=1}^{M}a_i W_{\gamma_i}^{1/2}(A) := \Phi_F \left[\sum \limits_{i=1}^{M}a_i W_{\gamma_i}^{r,1/2} \right]\ ,
\end{align}
for any finite set of $M$ arbitrary loops $\gamma_i \in {\cal L}$ and $M$ complex numbers $a_i$.
The existence of this measure is proven by showing that it is continuous with respect to the $C^*$-norm on $\bar{\cal A}/\bar{\cal G}$. The proof is similar to the one in the Abelian case \cite{Var2} and it goes as follows.\\

The $C^*$-norm on $\bar{\cal A}/\bar{\cal G}$ is defined as
\begin{align}
\norm{\sum \limits_{i=1}^{M}a_i W_{\gamma_i}^{1/2}} := \sup \limits_{A \in \bar{\cal A}/\bar{\cal G}} \abs{\sum \limits_{i=1}^{M}a_i W_{\gamma_i}^{1/2}(A)} = \sup \limits_{A \in {\cal A}} \abs{\sum \limits_{i=1}^{M}a_i W_{\gamma_i}^{1/2}(A)}\ .
\end{align}
The positive linear functional $\Phi_F$ is defined in \eqref{FPLF} via the Fock vacuum expectation values of the r-Wilson operators. Using the standard Fock representation, the Fock space is identified as $L^2({\cal S}^*, d\nu^g)$ where ${\cal S}^*$ is an appropriate space of tempered distributions and $\nu^g$ is the standard Gaussian measure with unit volume. We therefore have that
\begin{align}
	\mean{\sum \limits_{i=1}^{M}a_i \hat{W}_{\gamma_i}^{r,1/2}} = \int_{A \in {\cal S}^*} d\nu^g\  \sum \limits_{i=1}^{M}a_i \hat{W}_{\gamma_i}^{r,1/2}(A)\ ,
\end{align}
which implies that
\begin{align}
\abs{\mean{\sum \limits_{i=1}^{M}a_i \hat{W}_{\gamma_i}^{r,1/2}}} \leq \sup \limits_{A \in {\cal S}^*} \abs{\sum \limits_{i=1}^{M}a_i W_{\gamma_i}^{r,1/2}(A)}\ .
\end{align}
Given that the smearing function $f_r$ is a Schwartz function, the smeared holonomies $h_\gamma^{r,J}(A)$ in \eqref{SU(2)_smeared_hol} are well defined $SU(2)$ holonomies of the connection $A^r$. This means that an element $A \in {\cal S}^*$ is mapped via the smearing \eqref{Ar} to a Lie algebra valued 1-form $A^r \in {\cal S}^* \cap \bar{\cal A}$. Hence we have that
\begin{align}
	\forall A \in {\cal S}^*,\ \exists A^r \in {\cal S}^* \cap \bar{\cal A}:\ \sum \limits_{i=1}^{M}a_i W_{\gamma_i}^{r,1/2}(A) = \sum \limits_{i=1}^{M}a_i W_{\gamma_i}^{1/2}(A^r)\ ,
\end{align}
and consequently
\begin{align}
\sup \limits_{A \in {\cal S}^*} \abs{\sum \limits_{i=1}^{M}a_i W_{\gamma_i}^{r,1/2}(A)} \leq \sup \limits_{A \in {\cal S}^*\cap \bar{\cal A}} \abs{\sum \limits_{i=1}^{M}a_i W_{\gamma_i}^{1/2}(A)} \leq \sup \limits_{A \in \bar{\cal A}} \abs{\sum \limits_{i=1}^{M}a_i W_{\gamma_i}^{1/2}(A)}\ .
\end{align}
We therefore have
\begin{align}
\abs{\Phi_F\left[ \sum \limits_{i=1}^{M}a_i W_{\gamma_i}^{r,1/2} \right]} = \abs{\mean{\sum \limits_{i=1}^{M}a_i \hat{W}_{\gamma_i}^{r,1/2}}} \leq \sup \limits_{A \in \bar{\cal A}/\bar{\cal G}} \abs{\sum \limits_{i=1}^{M}a_i W_{\gamma_i}^{r,1/2}(A)}\ ,
\end{align}
which concludes the proof of existence of the measure $\mu^r_{SU(2)}$ on $\bar{\cal A}/\bar{\cal G}$.\\

By virtue of \eqref{GaugeInvMeasure}, and the fact that the Wilson loops are gauge invariant, the measure $\mu^r_{SU(2)}$ can be extended to a gauge invariant measure on the whole space $\bar{\cal A}$, therefore defining the $L^2(\bar{\cal A},\mu_{SU(2)}^r)$ Hilbert space.

Having now defined the measure $\mu^r$ on the space $\bar{\cal A}$, one could ask about the relation between this r-Fock measure and the natural measure $\mu^o$. This relation could be established as follows.
Given the positive linear functional $\Phi_F$ on ${\cal{HA}}_r$, we can define a positive linear functional $\Phi_F^r$ on ${\cal{HA}}$ as
\begin{align}
\Phi_F^r \left[\sum \limits_{i=1}^{M}a_i W_{\gamma_i}^{1/2} \right] := \Phi_F \left[\sum \limits_{i=1}^{M}a_i W_{\gamma_i}^{r,1/2} \right]\ ,
\end{align}
for any finite set of $M$ arbitrary loops $\gamma_i \in {\cal L}$ and $M$ complex numbers $a_i$.

Recall that the loop Hilbert space admits the decomposition \eqref{Decomp} and an orthonormal basis of which the elements are called spin networks (in the case of $SU(2)$). Given a graph $\Gamma$, the measures $\mu^r$ and $\mu^o$ can be restricted to a subspace ${\bar{\cal A}}_\Gamma$, which is obtained by restricting the action of generalized connections in $\bar{\cal A}$ to the edges of the graph $\Gamma$, and we denote their restrictions $\mu^r_\Gamma$ and $\mu^o_\Gamma$ respectively. 

A spin network state $\Psi_{\Gamma,\{j,\iota\}} \in L^2({\bar{\cal A}}_\Gamma, \mu^o)$, can always be expressed in a non-unique way in terms of a finite linear combination of Wilson loops associated to a set of closed loops within the graph $\Gamma$. We then introduce a map $\mathcal{I}_\Gamma$ which associates to each spin network state in $L^2({\bar{\cal A}}_\Gamma, \mu^o)$ a finite set of loops and complex numbers $\{(\gamma_i, a_i)\}$ such that
\begin{align}
	\text{if}\ \mathcal{I}_\Gamma \left(\Psi_{\Gamma,\{j,\iota\}} \right) = \{(\gamma_1,a_1),\dots, (\gamma_M,a_M) \},\ \text{then}\ \forall A\in {\bar{\cal A}}_\Gamma,\ \Psi_{\Gamma,\{j,\iota\}}(A) = \sum \limits_{i=1}^{M}a_i W_{\gamma_i}^{1/2}(A)\ .
\end{align}
We therefore can write that for every spin network state $\Psi_{\Gamma,\{j,\iota\}}$ we have
\begin{align}
\Phi_F^r \left[\Psi_{\Gamma,\{j,\iota\}} \right] = 
\Phi_F^r \left[\sum \limits_{(\gamma_i,a_i)\in \mathcal{I}_\Gamma \left(\Psi_{\Gamma,\{j,\iota\}} \right)}a_i W_{\gamma_i}^{1/2} \right] = \Phi_F \left[\sum \limits_{(\gamma_i,a_i)\in \mathcal{I}_\Gamma \left(\Psi_{\Gamma,\{j,\iota\}} \right)}a_i W_{\gamma_i}^{r,1/2} \right]\ .
\end{align}
Upon a choice of intertwiner bases $\{\iota\}$, the spin networks form an orthonormal basis in $L^2({\bar{\cal A}}_\Gamma, \mu^o)$. We hence obtain
\begin{align}
	d\mu^r_\Gamma = \left(\sum \limits_{\{j,\iota\}_\Gamma} \Phi_F \left[\sum \limits_{(\gamma_i,a_i)\in \mathcal{I}_\Gamma \left(\Psi_{\Gamma,\{j,\iota\}} \right)}a_i W_{\gamma_i}^{r,1/2} \right] \overline{\Psi_{\Gamma,\{j,\iota\}}} \right) d\mu^o_\Gamma\ ,
\end{align}
and consequently
\begin{align}\label{rFockToAL}
d\mu_{SU(2)}^r = \left(\sum \limits_{\Gamma}\sum \limits_{\{j,\iota\}_\Gamma} \Phi_F \left[\sum \limits_{(\gamma_i,a_i)\in \mathcal{I}_\Gamma \left(\Psi_{\Gamma,\{j,\iota\}} \right)}a_i W_{\gamma_i}^{r,1/2} \right] \overline{\Psi_{\Gamma,\{j,\iota\}}} \right) d\mu_{SU(2)}^o\ .
\end{align}

\subsection{Generalization to $SU(N)$ gauge theory}

Based on the $SU(2)$ case, the generalization of the construction of a r-Fock measure to an arbitrary $SU(N)$ gauge theory is rather straightforward. Except for a few modifications in the calculation of the Fock vacuum expectation value of the r-Wilson loop operators, the steps and results of the construction are similar to the $SU(2)$ case. In the following we outline the main steps and results in the general case.

Given a $SU(N)$ gauge theory with a phase space parametrized by a $\mathfrak{su}(N)$ Lie algebra valued connection $A$ and the conjugate field $E$, the definitions of the smeared connection \eqref{Ar}, the smeared holonomies \eqref{SU(2)_smeared_hol} as well as their Fock quantized counter-parts \eqref{Aop} and \eqref{SU2-r-hol} are the same, up to the replacement of the $SU(2)$ generators $\tau_i$ by the appropriate $SU(N)$ generators, which we denote $\lambda_i$. Consequently, the r-Wilson loop operator in the $SU(N)$ case takes the form
\begin{align}\label{SU(N)-r-Wilson}
\hat{W}_\gamma^{r,J} := \Tr \left[ \Pexp \left[ \int_\gamma ds \int \frac{d^3k}{q\sqrt{2|k|}} \lambda_i^{(J)}  \tilde{X}^a_{\gamma,r}(s,k) \left( c_a^{i \dag} (k) + c_a^i (-k) \right) \right] \right]\ ,
\end{align}
where we kept the same notation for the canonical ladder operators and the group representation label $J$ for the $SU(N)$ generators.

We are therefore interested in calculating the Fock vacuum expectation value
\begin{align}
 \mean{\hat W_{\gamma_1}^{r,J} \dots \hat W_{\gamma_{N-1}}^{r,J} }\ ,
\end{align}
for an arbitrary set of loops $\{\gamma_1,\dots,\gamma_{N-1}\} \in {\cal{L}}^{N-1}$.
Similarly to the $SU(2)$ case, the calculation of this expectation value is to be done in the fundamental representation, denoted $j_o$, of the $SU(N)$ group under consideration. 
Using the expansion of the r-Wilson operators as in \eqref{ExpansionWr}, then performing the product of $N-1$ such operators and rearranging the results as a single series, one obtains
{\small
	\begin{align}\label{WrProduct}
	\hspace{-0,3cm}
	 \mean{\hat W_{\gamma_1}^{r,j_o} \dots \hat W_{\gamma_{N-1}}^{r,j_o} }
	= \sum \limits_{n=0}^\infty  & \left( \phantom{\rule[-20pt]{1pt}{50pt}}
	\sum \limits_{p_1=0}^{2n} \dots \sum \limits_{p_{N-2}=0}^{2n-\sum \limits_{k=1}^{N-3} p_k}
	 \Tr \left[ \prod \limits_{m_1=1}^{p_1} \lambda^{j_o}_{i_{m_1}} \right]  \dots  \Tr \left[ \prod \limits_{m_{N-2}=\sum \limits_{k=1}^{N-3} p_k+1}^{\sum \limits_{k=1}^{N-2} p_k} \lambda^{j_o}_{i_{m_{N-2}}} \right]  \Tr \left[ \prod \limits_{m_{N-1}=\sum \limits_{k=1}^{N-2} p_k + 1}^{2n-\sum \limits_{k=1}^{N-2} p_k} \lambda^{j_o}_{i_{m_{N-1}}} \right]      \right. \\
	 & \times \left. \left( \prod \limits_{t=1}^{N-1} \prod \limits_{m_t=p_{t-1}+1}^{p_t} \ \Pint{\gamma_t} ds_{m_t} \int \frac{d^3k_{m_t}}{q\sqrt{2|k_{m_t}|}} \tilde{X}^{a_{m_t}}_{\gamma_t,r}(s_{m_t},k_{m_t}) \mean{ \prod \limits_{m=1}^{2n} \left( c_{a_m}^{i_m \dag} (k_m) + c_{a_m}^{i_m} (-k_m) \right)} \right) \phantom{\rule[-20pt]{1pt}{50pt}} \right)\ .
	\end{align}
}
One can see that the difference in the calculation of the final result with respect to $SU(2)$ is in the presence of a product of traces of a product of $SU(N)$ generators. Unlike the rather simple coefficients $\Upsilon_{\sigma^{(2n)}}^{(1/2)}$ in \eqref{Y2n} for $SU(2)$, estimating the trace of a product of generators for an arbitrary $SU(N)$ group as well as performing the contraction of the algebra indices over a product of traces is a much more complicated task. Nevertheless, the expectation value \eqref{WrProduct} is well defined and one can prove the convergence of the series (see appendix \ref{AppendixA}). 

The expectation value \eqref{WrProduct} allows to introduce the positive linear functional $\Phi_F$ on the $SU(N)$ smeared holonomies algebra as in \eqref{FExpWr}. The proof of positivity of the functional $\Phi_F$ follows the same reasoning as in the $SU(2)$ case and relies simply on the validity of the $SU(N)$ Mandelstam identities for the smeared Wilson loops and the r-Wilson loop operators. One then defines the r-Fock measure $\mu_{SU(N)}^r$ on ${\bar{\cal A}/\bar{\cal G}}$ as in \eqref{rFockMeasure}, with the proof of existence being the same as in the $SU(2)$ case, since the key point in the proof is the fact that the smearing of a connection $A\in {\cal S}^*$ using a Schwartz function produces a connection $A^r \in {\cal S}^* \cap \bar{\cal A}$. Finally, the r-Fock measure is extended to the whole space $\bar{\cal A}$ by virtue of \eqref{GaugeInvMeasure}, and one obtains a similar relation between the r-Fock measure and the natural measure as in \eqref{rFockToAL}.

\section{Summary \& comments}

In the present article, we introduced r-Fock measures for $SU(N)$ gauge theories on the space $\bar{\cal A}$ of generalized connections, that is the configuration space of the loop quantum theory, and hence generalizing the earlier works of M.Varadarajan for Abelian gauge theories \cite{Var1, Var2, Var3}. After recalling the construction of the r-Fock reprentation for the Abelian gauge theory, we focused on the definition of an r-Fock measure for an $SU(2)$ gauge theory as a concrete example. We first presented the algebra of $SU(2)$ smeared holonomies around closed loops ${\cal{HA}}_r$ on the three dimensional Riemannian flat space, then we moved to the standard Fock representation and performed the calculation of the Fock vacuum expectation value of the trace of an arbitrary smeared holonomy operator, which we call the r-Wilson loop operator. The result of this calculation is used to define a positive linear functional $\Phi_F$ on the algebra of smeared holonomies. We then use the functional $\Phi_F$ to define a gauge invariant measure $\mu_{SU(2)}^r$ on the space $\bar{\cal A}$, which is the r-Fock measure for the $SU(2)$ gauge theory, and expose the relation between the new measure and the natural difeomorphism invariant measure $\mu_{SU(2)}^o$. Finally, we present how the construction extends to arbitrary $SU(N)$ gauge group to obtain the corresponding r-Fock measure.

It is important to recall few aspects which display the contrast between the Abelian case and the non Abelian case. Unlike the Abelian case, in which the smeared holonomies-electric field algebra is isomorphic to the standard holonomy-flux algebra, the two algebras for a non Abelian $SU(N)$ gauge group are not. Therefore, in the non Abelian case, the Fock representation of the smeared holonomies algebra ${\cal{HA}}_r$ is not a representation of the standard holonomy-flux algebra. An aspect of this disparity is that the smeared electric field is not a well defined operator on the space of cylindrical functions. Furthermore, it is not clear yet whether the Fock representation of the smeared holonomies algebra ${\cal{HA}}_r$ is unitarily equivalent to the standard Fock representation of the connection algebra. The relation between these two representations is important for the construction of shadow states in the loop theory space, and their interpretation. This is a work in progress.

\subsection*{Acknowledgment}

This work was supported by the Polish National Science Center OPUS 15 Grant No.~2018/29/B/ST2/01250.


\appendix

\section{Convergence of the result of the expectation value of r-Wilson loop operators}\label{AppendixA}

i) Proof that the series in \eqref{FExpWr} is absolutely convergent:\\

From \eqref{FExpWr} we have
\begin{align}
 \abs{\mean{\hat{W}_\gamma^{r,1/2}}}
 &= \abs{ \sum \limits_{n=0}^\infty \frac{1}{2^n n!}\Pint{\gamma} ds_1 \dots ds_{2n} \sum_\sigma \Upsilon_{\sigma^{(2n)}}^{(1/2)} \left(\prod \limits_{m=1}^{n} \int \frac{d^3k}{2 q^2 |k|} \tilde{X}^{a}_{\gamma,r}(s_{\sigma(2m-1)},k) \tilde{X}^{a}_{\gamma,r}(s_{\sigma(2m)},-k) \right) } \\ 
 &= \abs{ \sum \limits_{n=0}^\infty \frac{1}{2^n n!} \Pint{\gamma} ds_1 \dots ds_{2n}\, \tilde{X}^{(2)}_{\gamma,r}(s_1,\dots,s_{2n})}\\
 &\leq \sum \limits_{n=0}^\infty \frac{1}{2^n n!} \abs{\Pint{\gamma} ds_1 \dots ds_{2n}\, \tilde{X}^{(2)}_{\gamma,r}(s_1,\dots,s_{2n})}\ .
\end{align}
where
\begin{align}
  \tilde{X}^{(2)}_{\gamma,r}(s_1,\dots,s_{2n}) :=  \sum_\sigma \Upsilon_{\sigma^{(2n)}}^{(1/2)} \left(\prod \limits_{m=1}^{n} \int \frac{d^3k}{2 q^2 |k|} \tilde{X}^{a}_{\gamma,r}(s_{\sigma(2m-1)},k) \tilde{X}^{a}_{\gamma,r}(s_{\sigma(2m)},-k) \right)\ .
\end{align}
Also
{\small
\begin{align}\label{Abs.Pint}
 & \abs{\Pint{\gamma} ds_1 \dots ds_{2n}\, \tilde{X}^{(2)}_{\gamma,r}(s_1,\dots,s_{2n})} \\ 
 & = \abs{\Pint{\gamma} ds_1 \dots ds_{2n} \sum_\sigma \Upsilon_{\sigma^{(2n)}}^{(1/2)} \left(\prod \limits_{m=1}^{n} \int \frac{d^3k}{2 q^2 |k|} \tilde{X}^{a}_{\gamma,r}(s_{\sigma(2m-1)},k) \tilde{X}^{a}_{\gamma,r}(s_{\sigma(2m)},-k) \right)} \\
 & \leq\ \Pint{\gamma} ds_1 \dots ds_{2n} \sum_\sigma \abs{\Upsilon_{\sigma^{(2n)}}^{(1/2)}} \abs{ \left(\prod \limits_{m=1}^{n} \int \frac{d^3k}{2 q^2 |k|} \tilde{X}^{a}_{\gamma,r}(s_{\sigma(2m-1)},k) \tilde{X}^{a}_{\gamma,r}(s_{\sigma(2m)},-k) \right)}\\
 & \leq\ \frac{3^n}{2^{2n-1}}\ \Pint{\gamma} ds_1 \dots ds_{2n} \sum_\sigma \abs{\prod \limits_{m=1}^{n} \delta_{a_{\sigma(2m-1)} a_{\sigma(2m)}} \int \frac{d^3k}{2 q^2 |k|} \tilde{X}^{a_{\sigma(2m-1)}}_{\gamma,r}(s_{\sigma(2m-1)},k) \tilde{X}^{a_{\sigma(2m)}}_{\gamma,r}(s_{\sigma(2m)},-k)}\ ,
\end{align}
}
where we used the fact that $\abs{\Upsilon_{\sigma^{(2n)}}^{(1/2)}}\leq {3^n}/{2^{2n-1}}$.  
Using the expression for the functions $\tilde{X}^{a}_{\gamma,r}(s,k)$ in \eqref{X.func}, we have
\begin{align}
\abs{\int \frac{d^3k}{2 q^2 |k|} \tilde{X}^{a_i}_{\gamma,r}(s_i,k) \tilde{X}^{a_j}_{\gamma,r}(s_j,-k)}
& = \abs{\dot{\gamma}^{a_i}(s_i)} \abs{\dot{\gamma}^{a_j}(s_j)} \abs{\int \frac{d^3k}{2 q^2 |k|} e^{-i \vec{k}.(\vec{\gamma}(s_i)-\vec{\gamma}(s_j))}\ \tilde{f}_r(k) \tilde{f}_r(-k)}  \ ,
\end{align}
For the integral on the right-hand side to produce a well defined on the whole $\mathbbm{R}^3$ domain, which corresponds (up to a factor $\sqrt{2\pi}^{3/2}/(2q^2)$) to the inverse Fourier transform of the function $\tilde{f}_r(k) \overline{\tilde{f}_r(k)}/\abs{k}$, the function $\tilde{f}_r(k) \overline{\tilde{f}_r(k)}/\abs{k}$ must be integrable. This means that we need the function $\tilde{f}_r(k)/\sqrt{\abs{k}}$ to be square integrable, which restricts the choice of the smearing function $f_r$. Assuming this condition, we further obtain
\begin{align}
\abs{\int \frac{d^3k}{2 q^2 |k|} \tilde{X}^{a_i}_{\gamma,r}(s_i,k) \tilde{X}^{a_j}_{\gamma,r}(s_j,-k)}
& \leq \abs{\dot{\gamma}^{a_i}(s_i)} \abs{\dot{\gamma}^{a_j}(s_j)} \int \frac{d^3k}{2 q^2 |k|} \abs{e^{-i \vec{k}.(\vec{\gamma}(s_i)-\vec{\gamma}(s_j))}\  \tilde{f}_r(k) \tilde{f}_r(-k)}  \\
& = \abs{\dot{\gamma}^{a_i}(s_i)} \abs{\dot{\gamma}^{a_j}(s_j)} \int \frac{d^3k}{2 q^2 |k|} \  \tilde{f}_r(k) \tilde{f}_r(-k)\\
& =:  \abs{\dot{\gamma}^{a_i}(s_i)} \abs{\dot{\gamma}^{a_j}(s_j)} w^r\ ,
\end{align}
where $w^r:=\int \frac{d^3k}{2 q^2 |k|} \  \tilde{f}_r(k) \tilde{f}_r(-k)$ depends only on the smearing function $f_r$, and it corresponds to the square of the $L^2$-norm of the function $\tilde{f}_r(k)/\sqrt{2q^2\abs{k}}$. It then follows that
{\small
	\begin{align}
	& \Pint{\gamma} ds_1 \dots ds_{2n} \sum_\sigma \abs{\prod \limits_{m=1}^{n} \delta_{a_{\sigma(2m-1)} a_{\sigma(2m)}} \int \frac{d^3k}{2 q^2 |k|} \tilde{X}^{a_{\sigma(2m-1)}}_{\gamma,r}(s_{\sigma(2m-1)},k) \tilde{X}^{a_{\sigma(2m)}}_{\gamma,r}(s_{\sigma(2m)},-k)} \\
	& \leq \Pint{\gamma} ds_1 \dots ds_{2n} \sum_\sigma \left(\prod \limits_{m=1}^{n} \delta_{a_{\sigma(2m-1)} a_{\sigma(2m)}} \abs{\dot{\gamma}^{a_{\sigma(2m-1)}}(s_{\sigma(2m-1)})} \abs{\dot{\gamma}^{a_{\sigma(2m)}}(s_{\sigma(2m)})} w^r \right) \\
	& = (w^r)^n\ \Pint{\gamma} ds_1 \dots ds_{2n} \abs{\dot{\gamma}^{a_1}(s_{1})} \dots  \abs{\dot{\gamma}^{a_{2n}}(s_{2n})}\ \sum_\sigma \left(\prod \limits_{m=1}^{n} \delta_{a_{\sigma(2m-1)} a_{\sigma(2m)}} \right) \\
	& = (w^r)^n \left( \int_{\gamma} ds_1 \int_{\gamma} ds_2 \abs{\dot{\gamma}^{a}(s_{1})} \abs{\dot{\gamma}^{a}(s_{2})} \right)^n \leq (w^r)^n \ ,
	\end{align}
}
where in the last line we used the fact that the tangent vectors are normalized as $\dot \gamma^a(s) \dot \gamma^a(s) = 1$. Therefore
\begin{align}
\sum \limits_{n=0}^\infty \frac{1}{2^n n!} \abs{\Pint{\gamma} ds_1 \dots ds_{2n}\, \tilde{X}^{(2)}_{\gamma,r}(s_1,\dots,s_{2n})} \leq \sum \limits_{n=0}^\infty \frac{1}{2^n n!} \frac{3^n}{2^{2n-1}} (w^r)^n = 2 e^{\frac{3}{8} w_\gamma^{r}}\ ,
\end{align}
which shows that the series in \eqref{FExpWr} converges absolutely.\\

ii) Elements of the proof that the series in \eqref{WrProduct} for a $SU(N)$ group is absolutely convergent:\\

The steps to show that the series in \eqref{WrProduct} converges absolutely are the same as in the proof for $SU(2)$ above. The main technical difference is in establishing a bound for the coefficients in \eqref{WrProduct} which depend on the traces of the products of $SU(N)$ generators. In the case of $SU(2)$, we used the fact that we have an expression for the coefficients $\Upsilon_{\sigma^{(2n)}}^{(1/2)}$ to estimate a bound. In the general $SU(N)$ case ($N\geq 2$), we can use generic properties of the trace and of the $SU(N)$ generators.

It follows from \eqref{WrProduct} that the coefficients for which we need to estimate a bound are
\begin{align}
 \hspace{-0,3cm}
 \zeta_{\sigma^{(2n)}}^{(j_o)} := & \delta^{i_{\sigma(1)}i_{\sigma(2)}} \dots \delta^{i_{\sigma(2n-1)}i_{\sigma(2n)}} 
 \Tr \left[ \prod \limits_{m_1=1}^{p_1} \lambda^{j_o}_{i_{m_1}} \right] \Tr \left[ \prod \limits_{m_2=p_1+1}^{p_1+p_2} \lambda^{j_o}_{i_{m_2}} \right]  \dots  
 \Tr \left[ \prod \limits_{m_{N-1}=\sum \limits_{k=1}^{N-2} p_k + 1}^{2n-\sum \limits_{k=1}^{N-2} p_k} \lambda^{j_o}_{i_{m_{N-1}}} \right] .
\end{align}
In the fundamental representation $j_o$, we take the $SU(N)$ generators to be $N\times N$ complex matrices satisfying
\begin{align}\label{Generators}
 \left[ \lambda_i , \lambda_j \right] = \sum \limits_{k=1}^{N^2-1}\varepsilon_{ijk} \lambda_k \quad , \quad \left\{ \lambda_i , \lambda_j \right\}  = -\frac{1}{N} \delta_{ij} \mathbbm{1}^{j_o} +  \sum \limits_{k=1}^{N^2-1}\vartheta_{ijk} \lambda_k\ ,
\end{align} 
where $\mathbbm{1}^{j_o}$ is the identity element on the $SU(N)$ group in representation $j_o$, $\varepsilon_{ijk}$ is a totally antisymmetric real valued tensor and corresponds to the structure constants of the group, and $\vartheta_{ijk}$ is a totally symmetric tensor. These tensors are determined by computing the traces of the product of the commutator or the anti-commutator in \eqref{Generators} with a single $\lambda_m$.

Given a complex $N\times N$ matrix $T$, the singular values $\varsigma_t$ of $T$ are positive numbers defined as the eigenvalues of the matrix $\sqrt{T^*T}$, where ${}^*$ denotes the adjoint matrix, ordered in a decreasing order with respect to the index $t$ ($1\leq t \leq N$), that is $\varsigma_1 \geq \varsigma_2 \geq \dots \varsigma_N$. These singular values satisfy several properties, in particular we have
\begin{align}
 &\forall\ T, S \in M_N(\mathbb{C}),\ \forall\ t\in \{1,\dots,N\}: \\ 
 & \abs{\Tr(T)} \leq \sum \limits_{t=1}^N \varsigma_t(T)\  ,\quad \varsigma_t(T) \leq \sqrt{\Tr(T^* T)}\  , \quad \text{and}\quad 
 \varsigma_t(ST) \leq \varsigma_1(S) \varsigma_t(T) \leq \varsigma_1(S) \varsigma_1(T) \  .
\end{align}
It follows from these properties that for a product of $SU(N)$ generators $\lambda_i$ we have
\begin{align}
	\forall\ p\in \mathbb{N} \backslash \{0\}, \ 
	\abs{\Tr \left[ \prod \limits_{m=1}^{p} \lambda^{j_o}_{i_{m}} \right]} \leq  N \prod \limits_{m=1}^{p} \varsigma_1(\lambda^{j_o}_{i_{m}}) 
	\leq N \prod \limits_{m=1}^{p}  \sqrt{ \Tr\left([\lambda^{j_o}_{i_{m}}]^*\lambda^{j_o}_{i_{m}} \right)} = \frac{N}{2^{p/2}}\ ,
\end{align}
where the last equality follows from the properties in \eqref{Generators} and the fact that $[\lambda^{j_o}_{i_{m}}]^* = -\lambda^{j_o}_{i_{m}}$. Hence we obtain that
\begin{align}
\hspace{-0,3cm}
\delta^{i_{\sigma(1)}i_{\sigma(2)}} \dots \delta^{i_{\sigma(2n-1)}i_{\sigma(2n)}} 
\abs{\Tr \left[ \prod \limits_{m_1=1}^{p_1} \lambda^{j_o}_{i_{m_1}} \right]}  \dots   
\abs{\Tr \left[ \prod \limits_{m_{N-1}=\sum \limits_{k=1}^{N-2} p_k + 1}^{2n-\sum \limits_{k=1}^{N-2} p_k} \lambda^{j_o}_{i_{m_{N-1}}} \right]}
\leq N^{N-1} \frac{(N^2-1)^n}{2^n} \ ,
\end{align}
for every permutation $\sigma$ of $2n$ indices. The factor $(N^2-1)^n$ emerges from the contraction of the algebra indices via Kronecker deltas, where each delta stands for a sum of $N^2-1$ terms. Consequently, we get
\begin{align}
	\abs{\zeta_{\sigma^{(2n)}}^{(j_o)}} \leq N^{N-1} \frac{(N^2-1)^n}{2^n} \ .
\end{align}
We then have
{\small
	\begin{align}
	\nonumber
	\hspace{-0,3cm}
	\mean{\hat W_{\gamma_1}^{r,j_o} \dots \hat W_{\gamma_{N-1}}^{r,j_o} }
	& \leq \sum \limits_{n=0}^\infty  \left| \phantom{\rule[-20pt]{1pt}{50pt}}
	\sum \limits_{p_1=0}^{2n} \dots \sum \limits_{p_{N-2}=0}^{2n-\sum \limits_{k=1}^{N-3} p_k}
	\Tr \left[ \prod \limits_{m_1=1}^{p_1} \lambda^{j_o}_{i_{m_1}} \right]  \dots  \Tr \left[ \prod \limits_{m_{N-2}=\sum \limits_{k=1}^{N-3} p_k+1}^{\sum \limits_{k=1}^{N-2} p_k} \lambda^{j_o}_{i_{m_{N-2}}} \right]  \Tr \left[ \prod \limits_{m_{N-1}=\sum \limits_{k=1}^{N-2} p_k + 1}^{2n-\sum \limits_{k=1}^{N-2} p_k} \lambda^{j_o}_{i_{m_{N-1}}} \right] \right.\\
	& \qquad \quad  \times \left. \left(  \prod \limits_{t=1}^{N-1} \prod \limits_{m_t=p_{t-1}+1}^{p_t} \ \Pint{\gamma_t} ds_{m_t} \int \frac{d^3k_{m_t}}{q\sqrt{2|k_{m_t}|}} \tilde{X}^{a_{m_t}}_{\gamma_t,r}(s_{m_t},k_{m_t}) \mean{ \prod \limits_{m=1}^{2n} \left( c_{a_m}^{i_m \dag} (k_m) + c_{a_m}^{i_m} (-k_m) \right)} \right) \right| \\
	& \leq \sum \limits_{n=0}^\infty  N^{N-1} \frac{(N^2-1)^n}{2^n} \left| \phantom{\rule[-20pt]{1pt}{50pt}}  \sum \limits_{p_1=0}^{2n} \dots \sum \limits_{p_{N-2}=0}^{2n-\sum \limits_{k=1}^{N-3} p_k} \prod \limits_{t=1}^{N-1}\left(  \prod \limits_{m_t=p_{t-1}+1}^{p_t} \ \Pint{\gamma_t} ds_{m_t} \int \frac{d^3k_{m_t}}{q\sqrt{2|k_{m_t}|}} \tilde{X}^{a_{m_t}}_{\gamma_t,r}(s_{m_t},k_{m_t}) \right)
	  \right. \\
	 &\hspace{4cm}  \times \left. \phantom{\rule[-20pt]{1pt}{50pt}} 
	 \frac{1}{2^n n!}\sum_\sigma \prod \limits_{m=1}^{n} \delta_{a_{\sigma(2m-1)} a_{\sigma(2m)}} \delta^{(3)}(k_{\sigma(2m-1)}, k_{\sigma(2m)}) \right|\\
	 & \leq N^{N-1}  \sum \limits_{n=0}^\infty \frac{(N^2-1)^n}{2^{2n} n!} \sum \limits_{p_1=0}^{2n} \dots \sum \limits_{p_{N-2}=0}^{2n-\sum \limits_{k=1}^{N-3} p_k} (w^r)^n 
	 = N^{N-1}  \sum \limits_{n=0}^\infty \frac{(N^2-1)^n}{2^{2n} n!} (w^r)^n 
	 \left( 
	 \begin{array}{c}
	 n+N-2 \\ N-2
	 \end{array}
	 \right) \ .
	\end{align}
}
Performing the sum over $n$ leads to the final result
\begin{align}
	\mean{\hat W_{\gamma_1}^{r,j_o} \dots \hat W_{\gamma_{N-1}}^{r,j_o} }
	& \leq N^{N-1} {}_1F_1\left(N-1, 1, \frac{(N^2-1)w^r}{4}\right)\ ,
\end{align}
which shows that the series in \eqref{WrProduct} converges, and it is absolutely convergent.

\section{Mandelstam identities for $SU(2)$ smeared Wilson loops}\label{AppendixB}

The smeared Wilson loop is the trace of a smeared $SU(2)$ holonomy, namely
\begin{align}
 W_\gamma^{r,J} := \Tr \left[ h_\gamma^{r,J}(A) \right]\ .
\end{align}
Since
\begin{align}
 h_\gamma^{r,J}(A) = \Pexp \left[ \int_\gamma ds\ \dot{\gamma}^a(s) A_a^{r,i}(\gamma(s)) \tau_i^{(J)}  \right] \ ,
\end{align}
where $A_a^{r,i}(x):= \int_{R^3} d^3y f_r(x-y) A_a^i(y)$ are the components of the Lie algebra valued 1-form $A^r$, the smeared holonomies $h_\gamma^{r,J}(A)$ are holonomies for the connection $A^r$ and they satisfy
\begin{align}
 h_{\rho}^{r,J} = \mathbbm{1}^J \ , \quad h_{\gamma_1}^{r,J} h_{\gamma_2}^{r,J} = h_{\gamma_1 \circ \gamma_2}^{r,J}\ , \quad \left( h_{\gamma}^{r,J} \right)^{-1}= \left(h_{\gamma}^{r,J} \right)^* = h_{\gamma^{-1}}^{r,J} \ ,
\end{align}
where $\rho$ is the trivial loop (class), $\mathbbm{1}^J$ is the identity element on the $SU(2)$ group in the $J$ representation, and the ${}^*$ is the adjoint operation for the $SU(2)$ group component of the operator, not the adjoint map for operators on the Fock space denoted ${}^\dag$.
It follows that their traces satisfy the Mandelstam identities \cite{LQG0} for $SU(2)$, which means that the smeared Wilson loops also satisfy the Mandelstam identities. In particular we have
\begin{align}
 W_{\gamma_1 \circ \gamma_2}^{r,1/2} &= W_{\gamma_2 \circ \gamma_1}^{r,1/2} \ , \quad
 W_{\gamma_1}^{r,1/2}\ W_{\gamma_2}^{r,1/2} &= W_{\gamma_1 \circ \gamma_2}^{r,1/2} + W_{\gamma_1 \circ \gamma_2^{-1}}^{r,1/2}\ ,
\end{align}
which are the Mandelstam identities of the first and second kind respectively. We also have that
\begin{align}
 \forall \gamma \in \mathcal{L},\qquad \abs{W_{\gamma}^{r,1/2}} \leq W_{\rho}^{r,1/2} = 2 \ .
\end{align}
Using the last two equations it follows that
\begin{align}
 W_{\gamma}^{r,1/2} = W_{\gamma^{-1}}^{r,1/2} \ .
\end{align}

\section{Mandelstam identities for $SU(2)$ r-Wilson loop operators}\label{AppendixC}

The fact that the r-Wilson loop operators satisfy a sort of operator counterparts of the classical Mandelstam identities is intuitive, but not straightforward because we are dealing with operators on the Fock space. Hence, one needs to review the derivation of the Mandelstam identities in order to make sure that the promotion of the Wilson loops to operators on the Fock space does not spoil the properties they satisfy.

We begin with few important observations. Since the operators $\hat A_a^{r,i} (x)$ in \eqref{Aop} commute with each other, it follows that the $SU(2)$ smeared holonomy operators in \eqref{SU2-r-hol} satisfy
\begin{align}
 \hat h_{\rho}^{r,J} = \mathbbm{1}^J \otimes \hat{\mathbb{I}} \quad \text{and} \quad \hat h_{\gamma_1}^{r,J} \hat h_{\gamma_2}^{r,J} = \hat h_{\gamma_1 \circ \gamma_2}^{r,J} \ ,
\end{align}
where $\hat{\mathbb{I}}$ is the identity operator on the Fock space. This implies that
\begin{align}\label{B3}
 \hat h_{\gamma}^{r,J} \hat h_{\gamma^{-1}}^{r,J} = \mathbbm{1}^J \otimes \hat{\mathbb{I}} \quad \text{and} \quad \left( \hat h_{\gamma}^{r,J} \right)^{-1}= \left(\hat h_{\gamma}^{r,J} \right)^* = \hat h_{\gamma^{-1}}^{r,J} \ .
\end{align}
Consequently, we have
\begin{align}
 \hat W_{\rho}^{r,J} = (2J+1)\,\hat{\mathbb{I}} \quad \text{and} \quad \det{}_G\left[{\hat h_{\rho}^{r,J}} \right] = \hat{\mathbb{I}} \ ,
\end{align}
where $\det{}_G$ stands for the determinant of the group component of the operator. Then, thanks to the fact that the operators $\hat A_a^{r,i} (x)$ are self-adjoint, it follows that
\begin{align}
 \forall \gamma \in \mathcal{L},\ \det{}_G\left[{\hat h_{\gamma}^{r,J}} \right] = \det{}_G\left[{\hat h_{\rho}^{r,J}} \right] = \hat{\mathbb{I}} \ ,
\end{align}
which consists of a generalization of the unit determinant property of the $SU(2)$ group elements.

One can now proceed with the derivation of the Mandelstam identities for the r-Wilson loop operators. The Mandelstam identities of the first kind follow from the cyclic property of the trace and the commutativity of the operators $\hat A_a^{r,i} (x)$, and we have
\begin{align}
 \hat W_{\gamma_1 \circ \gamma_2}^{r,J} = \hat W_{\gamma_2 \circ \gamma_1}^{r,J} \ .
\end{align}
The second family of Mandelstam identities is part of the identities of the second kind. These identities are derived from the fact that in $K$ dimensions, a $K+1$ totally anti-symmetric tensor identically vanishes. The contraction of the tensor $\delta_{[B_1}^{A_1}\dots \delta_{B_{K+1}]}^{A_{K+1}}$ with $K+1$ holonomies, or smeared holonomies in our case, gives rise to the second family of Mandelstam identities. Following the derivation developed in \cite{GamTri}, and using the results presented above, one arrives at the desired identities. A particularly important identity for our work is
\begin{align}\label{B7}
 \hat W_{\gamma_1}^{r,1/2}\ \hat W_{\gamma_2}^{r,1/2} &= \hat W_{\gamma_1 \circ \gamma_2}^{r,1/2} + \hat W_{\gamma_1 \circ \gamma_2^{-1}}^{r,1/2} \ ,
\end{align}
which we use to show that the linear functional defined in \eqref{FPLF} is a positive linear functional on the algebra of smeared Wilson loops. 

Finally, using the self-adjointness of the operators $\hat A_a^{r,i} (x)$ and eqs.\ \eqref{B3} and \eqref{B7}, we get that the r-Wilson loop operators $\hat W_{\gamma}^{r,J}$ are self-adjoint operators:
\begin{align}
 \left(\hat W_{\gamma}^{r,J} \right)^\dag = \hat W_{\gamma}^{r,J}\ .
\end{align}


\bibliographystyle{ieeetr}
\bibliography{references}

\begin{thebibliography}{10}

\bibitem{LQG0}
R.~Gambini and J.~Pullin, {\em {Loops, Knots, Gauge Theories and Quantum
  Gravity}}.
\newblock {Cambridge University Press}, {1996}.

\bibitem{LQG1}
T.~Thiemann, {\em {Modern canonical quantum general relativity}}.
\newblock Cambridge University Press, 2008.

\bibitem{LQG2}
C.~Rovelli, {\em {Quantum gravity}}.
\newblock Cambridge Monographs on Mathematical Physics, Cambridge University
  Press, 2004.

\bibitem{LQG3}
A.~Ashtekar and J.~Lewandowski, ``{Background independent quantum gravity: A
  Status report},'' {\em Class.~Quant.~Grav.}, vol.~21, p.~R53, 2004.

\bibitem{LQG4}
M.~Han, W.~Huang, and Y.~Ma, ``{Fundamental structure of loop quantum
  gravity},'' {\em Int. J. Mod. Phys.}, vol.~D16, pp.~1397--1474, 2007.

\bibitem{Var1}
M.~Varadarajan, ``{Fock representations from U(1) holonomy algebras},'' {\em
  Phys. Rev.}, vol.~D61, p.~104001, 2000.

\bibitem{Var2}
M.~Varadarajan, ``{Photons from quantized electric flux representations},''
  {\em Phys. Rev.}, vol.~D64, p.~104003, 2001.

\bibitem{Var3}
M.~Varadarajan, ``{Gravitons from a loop representation of linearized
  gravity},'' {\em Phys. Rev. D}, vol.~66, p.~024017, 2002.

\bibitem{AshLewSah}
A.~Ashtekar, J.~Lewandowski, and H.~Sahlmann, ``{Polymer and Fock
  representations for a scalar field},'' {\em Class. Quant. Grav.}, vol.~20,
  pp.~L11--1, 2003.

\bibitem{AshLew1}
A.~Ashtekar and J.~Lewandowski, ``{Relation between polymer and Fock
  excitations},'' {\em Class. Quant. Grav.}, vol.~18, pp.~L117--L128, 2001.

\bibitem{Baez}
J.~Baez, ``{Generalized measures in gauge theory},'' {\em Lett Math Phys},
  vol.~31, p.~213–223, 1994.

\bibitem{AshLew}
A.~Ashtekar and J.~Lewandowski, ``Differential geometry on the space of
  connections via graphs and projective limits,'' {\em Journal of Geometry and
  Physics}, vol.~17, no.~3, pp.~191--230, 1995.

\bibitem{ALmeasure}
A.~Ashtekar and J.~Lewandowski, ``{Representation theory of analytic holonomy
  C* algebras},'' in {\em Knots and Quantum Gravity} (J.~Baez, ed.), Oxford
  Lecture Series in Mathematics and its Applications 1, pp.~21--61, Oxford
  University Press, Oxford, 1994.

\bibitem{MarolfMourao}
D.~Marolf and J.~M. Mourao, ``{On the support of the Ashtekar-Lewandowski
  measure},'' {\em Commun. Math. Phys.}, vol.~170, pp.~583--606, 1995.

\bibitem{MeasuresGT}
A.~Ashtekar and J.~Lewandowski, ``Projective techniques and functional
  integration for gauge theories,'' {\em Journal of Mathematical Physics},
  vol.~36, no.~5, pp.~2170--2191, 1995.

\bibitem{GamTri}
R.~Gambini and A.~Trias, ``Gauge dynamics in the c-representation,'' {\em
  Nuclear Physics B}, vol.~278, no.~2, pp.~436--448, 1986.

\bibitem{Wick}
G.~C. Wick, ``The evaluation of the collision matrix,'' {\em Phys. Rev.},
  vol.~80, pp.~268--272, Oct 1950.

\end{thebibliography}

\end{document}